\documentclass[12pt]{article}
\usepackage[margin=1in]{geometry}

\usepackage{times}
\usepackage{graphicx}
\usepackage[font=small]{caption}
\usepackage{subcaption}
\usepackage{titlesec}
\usepackage{ragged2e}
\usepackage{setspace}
\usepackage{tabu}
\usepackage{geometry}
\usepackage{caption}
\usepackage{subcaption}
\usepackage{amsmath}
\usepackage{enumerate}
\usepackage{rotating}
\usepackage{float}
 \usepackage{url}
\graphicspath{ {./Images/} }			
\titleformat{\section}
  {\normalfont\normalsize}
  {\thesection}{1em}{}
\titleformat{\subsection}
  {\normalfont\small}
  {\thesubsection}{1em}{}

\title{\bf \large Effect of recommending users and opinions on the network connectivity and idea generation process}
\author{\normalsize Sriniwas Pandey$^{1*}$ and Hiroki Sayama$^{1,2}$\\
\small$^1$ Binghamton Center of Complex Systems, Binghamton University, Binghamton, USA\\
\small$^2$ Faculty of Commerce, Waseda University, Japan\\
\small$^*$ spandey4@binghamton.edu}

\date{}

\begin{document}

\maketitle

\thispagestyle{empty}

\begin{center}Abstract\end{center}

The growing reliance on online services underscores the crucial role of recommendation systems, especially on social media platforms seeking increased user engagement. This study investigates how recommendation systems influence the impact of personal behavioral traits on social network dynamics. It explores the interplay between homophily, users' openness to novel ideas, and recommendation-driven exposure to new opinions. Additionally, the research examines the impact of recommendation systems on the diversity of newly generated ideas, shedding light on the challenges and opportunities in designing effective systems that balance the exploration of new ideas with the risk of reinforcing biases or filtering valuable, unconventional concepts.

\section{Introduction}

\par With the increasing popularity of social networks, they have proven to be  a great tool to communicate opinions, spread the information and influence opinions \cite{ferguson2014social, zeitel2014social}. However, the spread of information and opinions over social networks is not unimpeded and free. It is controlled and moderated for several reasons and by several parties including but not limited to social media algorithms, political and business entities \cite{artime2020effectiveness, badawy2018analyzing, engesser2017populism}. We have seen a massive development in the field of recommendation systems over the last decades. Recommendation systems have emerged as essential tools for numerous industries, providing personalized and relevant recommendations to individual users, which is critical for generating revenue and gaining a competitive edge in e-commerce platforms \cite{ISINKAYE2015261,adomavicius2005toward}. Recommendation systems find application in a diverse range of fields, including e-commerce, entertainment, research, personalized health and fitness, and more \cite{ISINKAYE2015261, recommendationSurvey2023, aggarwal2016recommender}.

The core functionality of recommendation systems lies in their ability to collect user data and analyze it to generate customized recommendations. These systems rely on both implicit and explicit data, such as browsing history, past purchases, and ratings provided by the user \cite{ISINKAYE2015261, recommendationSurvey2023, aggarwal2016recommender}. Over the years, various types of recommendation systems have been developed, each with its own strengths and limitations, depending upon the application, efficiency, and evaluation techniques \cite{adomavicius2005toward, gedikli2014should, ricci2015recommender, koren2021advances}. One new popular approach of recommendation systems is hybrid recommender systems, which use multiple data sources to generate more robust inferences. In the parallel hybrid design, input is provided to multiple recommendation systems, and the output of each of these recommendations is combined to generate a final output \cite{ccano2017hybrid, burke2002hybrid}.

Besides the development of recommendation systems, several researchers have worked on other aspects of recommendation systems, such as evaluation, privacy issues, and the effects of recommendation systems on user behavior \cite{del2008evaluation, chen2017performance, breese2013empirical, KOTKOV2016180, lam2006you, omar2019towards, adomavicius2011recommender}. The ongoing research in this field aims to improve the accuracy and efficiency of recommendation systems while addressing privacy concerns and exploring the broader implications of recommendation systems on society.

Previous research has extensively studied the effect of recommender systems on social dynamics and their tendency to reinforce biases by promoting similar content and users \cite{terren2021echo, nikolov2015measuring, pariser2011filter, de2022modelling}. Many studies have confirmed the existence of echo chambers on social media platforms, which can lead to polarization of opinions and limited exposure to diverse viewpoints \cite{quattrociocchi2016echo, garimella2018political}. However, some studies have challenged this notion by showing that individuals behave differently depending on the topic and that the effect of echo chambers can be overestimated \cite{barbera2015tweeting}. 

Additionally, individuals' preferences play a significant role in shaping their exposure to different viewpoints, and social media users tend to access news from more sources than non-users \cite{bakshy2015exposure, fletcher2018people}. Geographical social networks have also been found to map onto online echo chambers, as demonstrated in the Brexit campaign on Twitter \cite{bastos2018geographic}. Furthermore, effect of homophilic clusters appears to be more dominant for echo chamber formation than the bias in information consumption towards similar minded individuals, with Facebook being more segregated than Reddit \cite{cinelli_echo_2021}. In terms of political conversations, the presence of echo chambers varies across different social networks, with Reddit showing weaker evidence of echo chambers compared to other platforms \cite{de2021no}. 

Another important area of research that has gained attention is the study of opinion dynamics in the presence of recommender systems and other algorithmic interventions. Sirbu et al. \cite{sirbu2019algorithmic} have used a modified bounded confidence model to demonstrate that algorithmic biases, introduced to increase social network platform usage, can lead to polarization and fragmentation of the network. Perra et al. \cite{perra2019modelling} have presented a model to analyze the effect of algorithmic personalization on polarization and opinion dynamics. Fabbri et al. \cite{fabbri2020effect} have investigated the role of homophily in deciding the visibility of minorities in a bi-populated network in the presence of recommender systems. Recently, Cinus et al. \cite{cinus2022effect} have proposed a framework to measure the effect of people recommender systems on opinion dynamics, specifically on polarization and echo chamber formation.

These studies contribute significantly to our understanding of the complex relationship between recommender systems and opinion dynamics in social networks. We extend one of the recent works of Sayama \cite{sayama_extreme_2020} by incorporating the concept of recommender systems in it. In his work, Sayama investigates the role of conformity, homophily and attention to novel opinions on the network dynamics utilizing a numerical simulation model. We modify the existing model to include several recommendation and filtering strategies, to generate and collect opinions and to measure the opinion eccentricity \cite{pandey2023} to better understand the effect of recommendation systems on network dynamics and idea generation process.  By including recommender systems in the simulation model, we can examine how these systems impact the network's polarization, fragmentation, and the diversity of ideas. Our research extends the current understanding of opinion dynamics by incorporating recommender systems, which have become an integral part of social networks. We aim to identify potential risks associated with recommendation and filtering strategies and inform the design of more responsible and effective algorithms.

\section{Model}
\subsection{Existing model}
This paper builds upon Sayama's existing model \cite{sayama_extreme_2020} to analyze opinion dynamics on an adaptive social network. Specifically, the study explores the impact of personal behavioral traits, including homophily, conformity, and attention to novelty, on network dynamics. Through a series of systematic parameter sweep numerical experiments, the author concludes that homophily and attention to novel ideas have contradictory effects on network dynamics. When the homophily trait is dominant, it leads to fragmentation within the society. In contrast, attention to diverse opinions brings society closer with sameness in their ideologies.

\subsection{Updated model}
We are expanding this model in several ways, most importantly by incorporating recommendation strategies to investigate their impact on behavioral traits and network dynamics. The model begins with a fully connected directed network consisting of $n$ nodes, each representing a user with an initial idea state. Contrary to the original model, where the idea state is a scalar, idea state of a user is represented by a $k$-dimensional vector ${X} = \{x_1, x_2, ..., x_k\}$, where each $x_i$ is initialized from a uniform distribution $(0,1)$ (in our study $k$ was set to 15). The weight of each edge represents the degree of influence when exposed to an opinion. The initial connection weights are distributed between 0 and 1 by (i) a uniform random distribution, and (ii) a power law distribution. Edge weights following power law were generated using the following equation 

\begin{equation}
x = \frac{(1-u){x_{min}}^{1-\alpha}-u {x_{max}}^{1-\alpha}}{1-\alpha}
\label{eq:power_law_gen}
\end{equation}

where $x_{max} = 1$, $x_{min} = 10^{-6}$, $\alpha = 3$, and $u$ is a uniform random number. $\alpha$ value was chosen as $3$ since it represents one end of the spectrum within the typical range of $\alpha$ values in social systems ( $2 \leq \alpha \leq 3 $). The goal of this study is to analyze effects of recommendation systems; uniform and power law distributions are selected to exhibit the robustness of our findings. To precisely model any particular social network, the value of $\alpha$ should be adjusted accordingly.

At each time step, a randomly selected node $j$ generates a new opinion by a random fluctuation ($\overrightarrow{\mathcal{E}}$) from its idea state ${X}_j$ as follows.
\begin{equation}
{O}_j= {X}_j+ \overrightarrow{\mathcal{E}}
\end{equation}

At each time step both node states and edge weights undergo changes as well. However, to minimize computational complexity, we update the idea state and node weight only after every 100 time steps. Node states and edge weights co-evolve over time according to a modified version of the co-evolution mechanism used in the original model. The co-evolution mechanism is driven by homophily, attention to novelty, conformity, and random fluctuation, as in the original model. However, in our model, exposure to new opinions or users is controlled by the recommendation strategy being used.  $\overline{X^{exposed}_i}$ is the weighted average of all the opinions node \( i \) is exposed to( \( \overline{X^{exposed}_i} = \frac{\sum_{k} w_{ik} \cdot X_{k}}{\sum_{k} w_{ik}} \), where \( X_{k} \) is one of the opinions user \( i \) is exposed to,  \( k \) is its author,  and \( w_{ik} \) is the weight of the edge from user \( k \) to user \( i \) ). $\overline{X^{recent}_j}$ is the average of recent 10 opinions generated by node $j$  ( \(\overline{X^{recent}_j} = \frac{\sum_{m} X_{m}}{N}\), where  \( X_{m} \) represents one of the recent opinions generated by node \( j \),  \( N \) is the number of recent opinions considered, which is 10 in our study).
 
The co-evolution takes place as follows:
\begin{equation}
\Delta X_i = c\  (\overline{X^{exposed}_i }- \ X_i\ )+\overrightarrow{\mathcal{E}}.
\end {equation}
\begin{equation}
\Delta w_{ji} = h\  F_h(X_i,\overline{X^{recent}_j})\ +\ a\  F_a(\overline{X^{exposed}_i}, \overline{X^{recent}_j})
\label{eq:weight_Change}
\end {equation}
Here $\overline{X^{exposed}_i}$ is the weighted average of all the opinions node $i$ is exposed to.  Weighted average is calculated using the weight of edge from author of an opinion to user $i$; ;
$c$, $h$, and $a$ are the parameters determining strength of social conformity, homophily, and attention to novelty, respectively. $\mathcal{E}$ represents a random fluctuation.  $F_h$ and  $F_a$ are functions determining the edge weight change as follows:
\begin{equation}
F_h(X_i,\overline{X^{recent}_j})=\theta_h\ -\ \left\lvert {X}_i\ -\ \overline{X^{recent}_j}\right\rvert 
\end{equation}
\begin{equation}
F_a(\overline{X^{exposed}_i},\overline{X^{recent}_j})= \left\lvert \overline{X^{exposed}_i}\  -\  \overline{X^{recent}_j}\right\rvert -\theta_a
\end{equation}
where $\theta_h$, $\theta_a$ are fixed threshold values.
\subsection{Recommendation strategies}
After generating 100 new opinions in the network, we update node states and edge weights. Users are then exposed to some of the new opinions and users according to one of the following strategies:
\begin{itemize}
\item \textbf{20 strongest connections (SC):} This strategy involves each user being exposed to opinions generated by their 20 most strongly connected neighbors. Each user also updates their edge weights with these 20 neighbors according to equation \ref{eq:weight_Change}. This serves as the baseline for our analysis and is a popular method used in several social networks where no other recommendation system is used. The algorithm displays the top posts made by neighbors, which can eventually strengthen or weaken connections.

\item \textbf {20 Nearest opinions (NO):} In this recommendation strategy, each user is exposed to 20 opinions from the pool of 100 new opinions generated in the whole network that are most similar to their idea state. Each user also updates the weight of their edge with the authors of these 20 opinions. Nearness is calculated using the euclidean distance or $L_2$ norm between opinion vectors or idea states according to the strategy. 

\item \textbf {20 Farthest opinions (FO):} This strategy recommends 20 opinions from the pool of 100 new opinions generated in the whole network that are most distant from user’s idea state. The idea state of the user is updated by the recommended opinions, and the user also updates their edge weights with the authors of the recommended opinions.

\item \textbf {20 Nearest users (NU):} Instead of recommending opinions, this strategy recommends 20 users from the pool of active users (who have opinions in the pool of 100 new opinions ) from whole network. Users are selected based on the similarity of their idea state to that of the user. Recommending users does not impact the idea state but only updates connection weights.

\item \textbf {20 Farthest users (FU):} Similar to the previous strategy, only users are recommended, and the respective edge weights are updated. However, in this strategy, active users whose idea state is farthest from the idea state of user $u$ are recommended to user $u$.

\item \textbf {20 Nearest opinions and users (NOU):} This strategy is a combination of strategies NO and NU. Each user is recommended 20 nearest opinions and nearest users, and their idea state is impacted by the recommended opinions. Edge weights are updated for the connections with the recommended users and authors of recommended opinions.

\end {itemize}
\section{Experiments}
 We ran two sets of simulations to investigate the role of recommendation systems. The parameter combination used for both sets of experiments are as follows:
\begin {itemize}
\item $Network\ weights \in \{ uniform(0,1),\	 power law(0,1) \footnote{(\textbf{Equation \ref{eq:power_law_gen}})}\}$
\item $(h,a) \in \{(0.3, 0.01),(0.01,0.3)\}$ 
\item $Recommendation\ Strategy \in \{SC, NO, NU, FU, NOU, FO\}$
\end {itemize}
Parameter $c$,  $\theta_h$, and $\theta_a$ values are fixed as 0.01, 0.1, 0.1 for all the simulation runs. Conformity parameter $c$ controls the rate of change of a user's idea state.  0.01 is relatively smaller value for conformity $c$; meaning the neighbors' average idea state will have a smaller impact on a user's idea state. $\theta_h$ and $\theta_a$ decide the cutoff distance between idea states to decide the significance of interaction.

\subsection{Network dynamics}
The first set of experiments we conduct is to examine the relationship between recommendation strategy and network evolution. In the first set, we conducted 15 simulation runs for each combination of the above parameters.  The outcome measures we calculated in these experiments are maximum modularity of network and maximum standard deviation of average community state. We run each simulation until there are 3,000 opinions in the network and after every 100 new opinions we calculate the modularity and standard deviation of average community idea states. The Louvain's method \cite{blondel2008fast} is used to identify network fragmentation and communities in the network. 

\subsection{Idea generation process}
In the second set of experiments, we investigate the impact of recommendation strategies on eccentric idea generation. For each parameter combination, we calculate eccentricity of each opinion generated in the network and the overall distribution of eccentricities is observed. Eccentricity is the measure of strangeness of an opinion relative to the social neighborhood \cite{pandey2023}. Eccentricity has been proven an important metric in idea generation process to understand the social impact \cite{pandey2023characterizing}. Since eccentricity calculation is computationally complex, we only ran 3 simulation runs for each combination of parameters. Eccentricity calculation is performed as follows. The norm or center of a user's knowledgebase is calculated by taking the weighted average of neighbors' recent opinions. Distance of a new opinion from the center is the eccentricity of the new opinion. We use $L^2$ norm to find the distance between opinion and the center. 

\section{Results}
\subsection{Effect of recommendation strategies on network dynamics}
Network fragmentation is a phenomenon where a network breaks into smaller, disconnected components due to the formation of clusters or communities of nodes with similar characteristics or behaviors. High modularity of a network means that it consists of clusters of nodes that are densely connected within the cluster and sparsely connected between different clusters. These clusters are often referred to as communities or modules, and they are characterized by having a high degree of similarity or homogeneity in terms of the attributes of their nodes, such as their opinions, behaviors, or interests.

\begin{figure}

    \begin{subfigure}{0.49\textwidth}
         \centering
         \includegraphics[width=\textwidth]{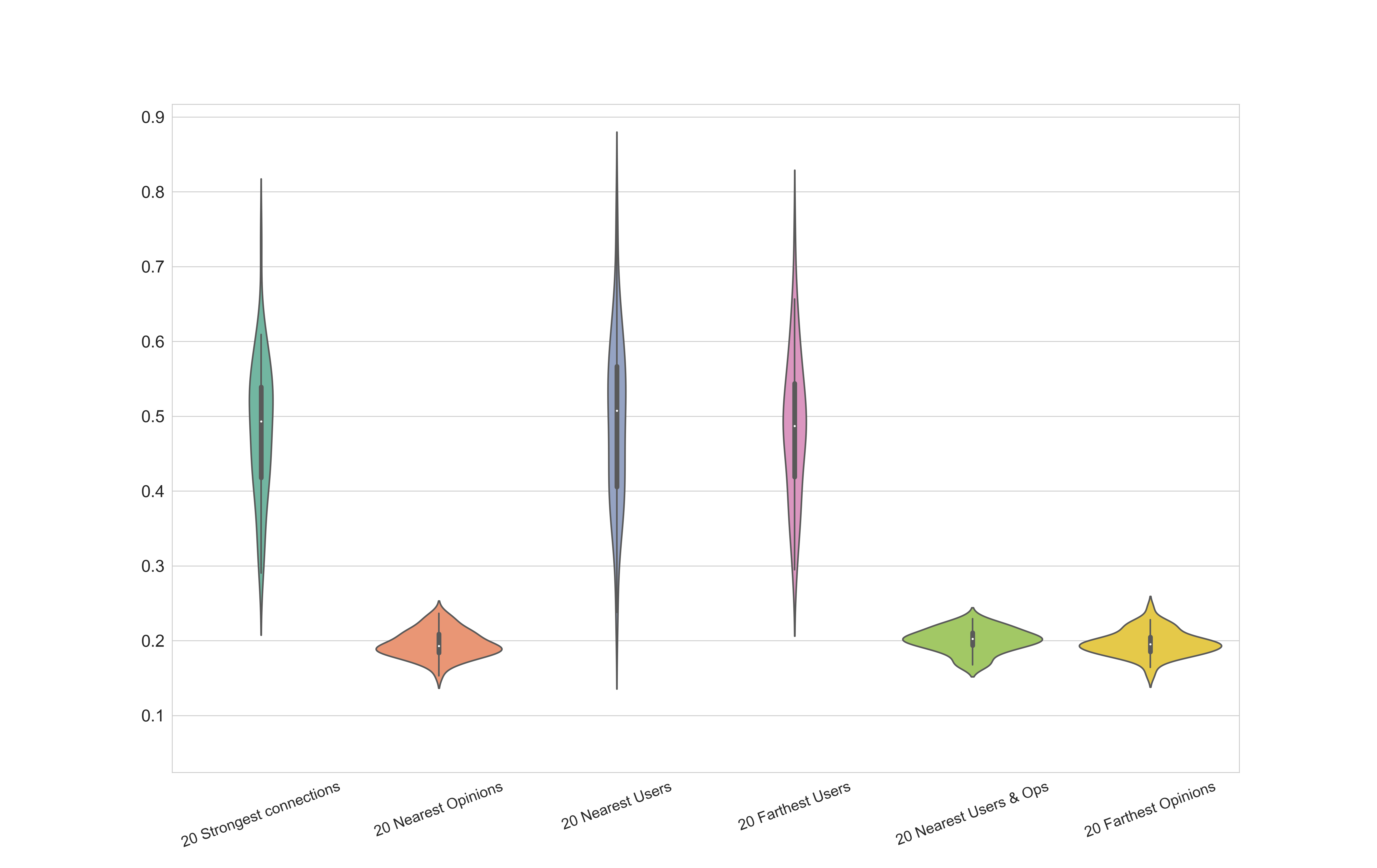}
         \caption{High homophily, low attention to novelty and initial weights follow power law distribution}
         \label{fig:modularity_violin_h3a01PL}
     \end{subfigure}
     \begin{subfigure}{0.49\textwidth}
         \centering
         \includegraphics[width=\textwidth]{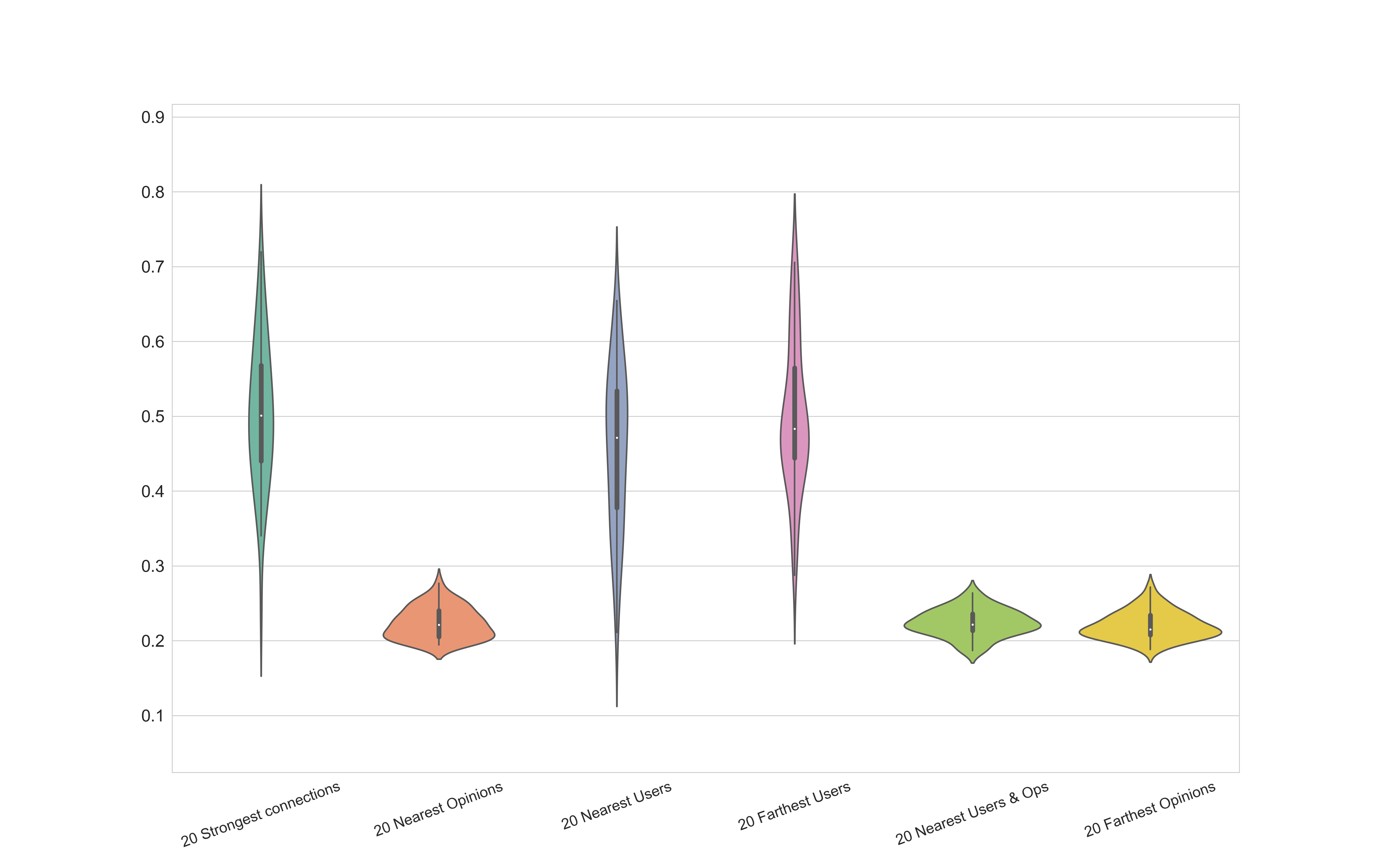}
         \caption{High homophily, low attention to novelty and initial weights follow uniform random distribution}
         \label{fig:modularity_violin_h3a01UD}
     \end{subfigure}
     \begin{subfigure}{0.49\textwidth}
         \centering
         \includegraphics[width=\textwidth]{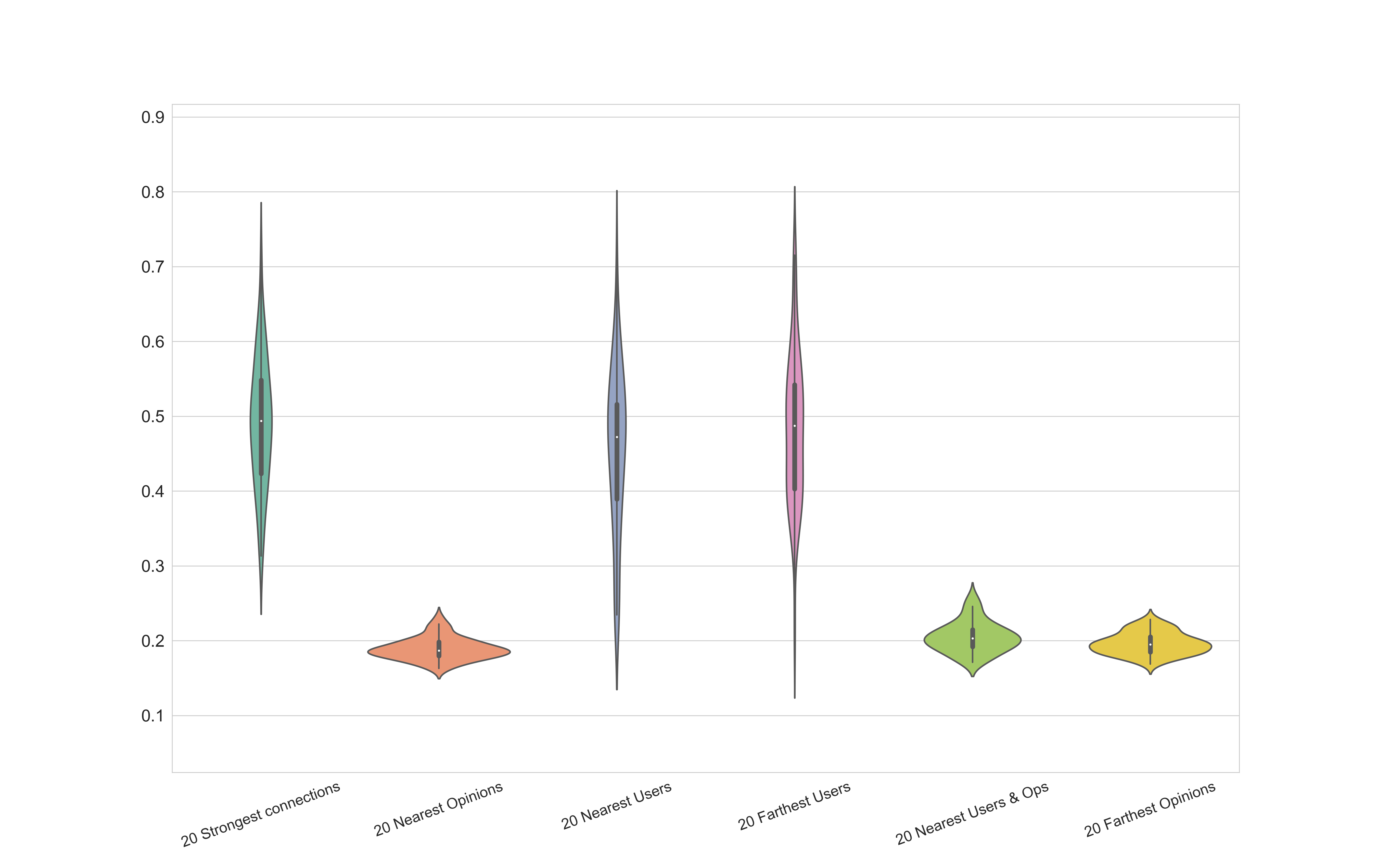}
         \caption{Low homophily, high attention to novelty and initial weights follow power law distribution}
         \label{fig:modularity_violin_h01a3PL}
     \end{subfigure}
     \begin{subfigure}{0.49\textwidth}
         \centering
         \includegraphics[width=\textwidth]{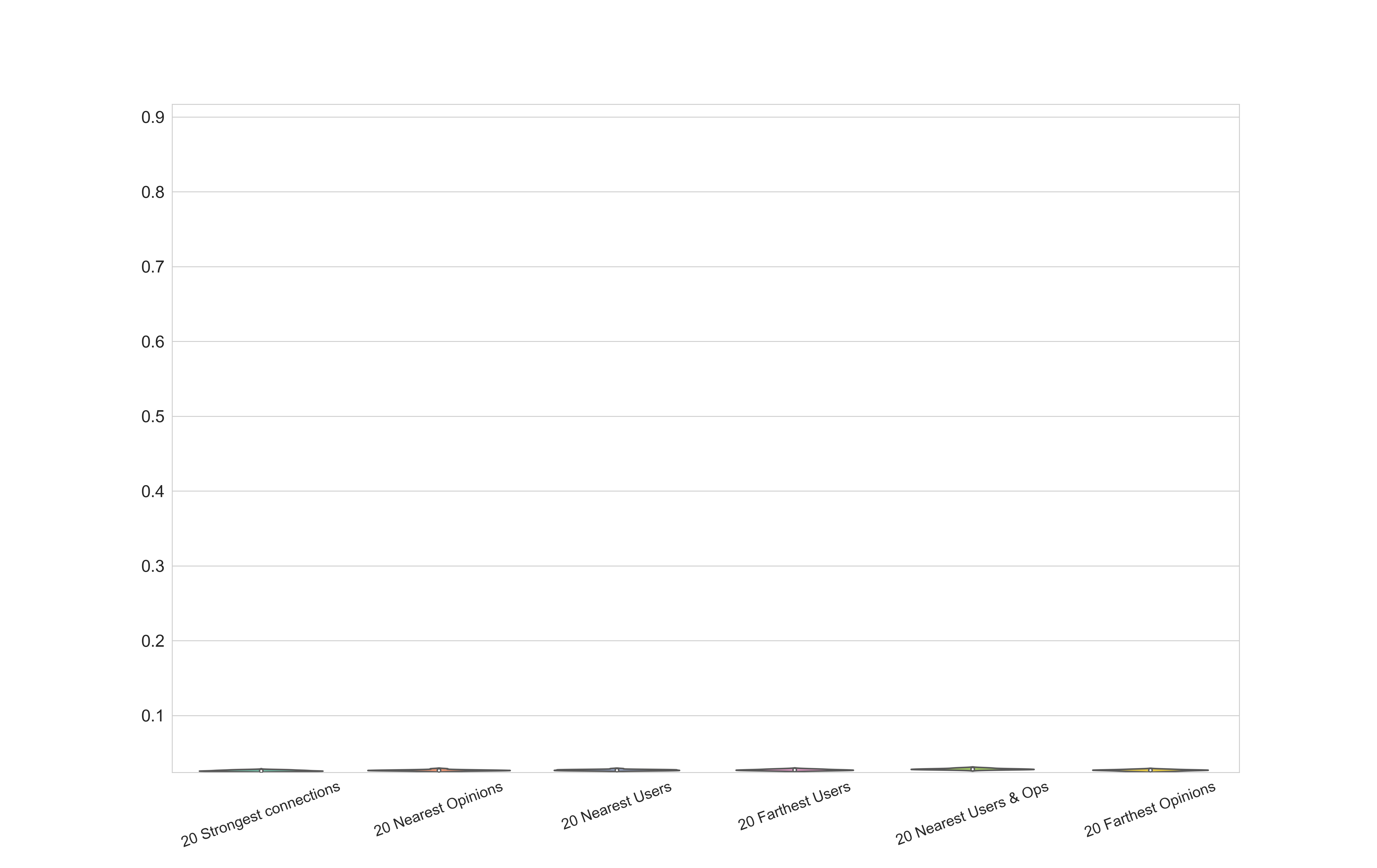}
         \caption{Low homophily, high attention to novelty and initial weights follow uniform random distribution}
         \label{fig:modularity_violin_h01a3UD}
     \end{subfigure}
     \caption{Violin plots representing distribution of network modularity for different recommendation strategies and parameter configurations: Each subfigure represents mudularities for the parameter setting described in the caption of subfigure. Modularity is calculated according to the partitions obtained using the Louvain community detection method.  In all the cases except (d) when homophily is low, attention to novelty is high and initial edge weights are distributed according to uniform random distribution, opinion recommendation has significantly lower modularity. }
\label{fig:modularity}
 \end{figure}

Our study shows that the use of opinion recommendation strategies (NO, FO) consistently discouraged network fragmentation across all scenarios when baseline strategy fragments the network. In the case of low homophily, high attention to novelty, and uniformly distributed weights, the network remains well connected and unfragmented even for the baseline strategy (SC) (and for all other strategies) (Figure \ref{fig:modularity}). This result implies that when users are exposed to diverse opinions and perspectives, they are more likely to maintain connections with others, which can contribute to a more cohesive network. This finding suggests that these strategies can help mitigate the negative effects of echo chambers and filter bubbles, where users only interact with like-minded individuals, leading to greater ideological polarization and network fragmentation.

We found that in baseline (SC) and user recommendation scenarios (NU, FU, NOU), the network modularity is significantly high, denoting the fragmentation of the network. This can be contributed to solely updating weights without adjusting the idea state of users. This observation highlights the importance of considering the underlying mechanisms of recommendation and filtering strategies when designing them to promote network cohesion.

Interestingly, in the case of low homophily and uniformly distributed weights, the network did not get fragmented regardless of the recommendation strategy used. This result suggests that in scenarios where there is a low tendency for users to connect with similar users and initial weights are uniform, the recommendation strategy may have less impact on network fragmentation. Nonetheless, it is essential to note that this scenario may not be representative of real-world social and information networks, where homophily and initial biases are prevalent. While opinion recommendation strategies can contribute to network cohesion, their effectiveness may depend on the underlying network structure and the characteristics of its nodes.

In social and information networks, communities are formed based on shared interests, beliefs, and preferences. These communities may have distinct ideological stances or opinions on different issues. The spread of communities in terms of their idea states can provide insight into the extent to which different groups within a network are ideologically aligned. To measure the spread of communities in terms of their idea states, we calculated the average idea state of nodes within a community and then computed the standard deviation of these average community idea states. A lower standard deviation indicates that the communities are more ideologically aligned, while a higher standard deviation indicates that the communities are more ideologically diverse.

\begin{figure}

    \begin{subfigure}{0.49\textwidth}
         \centering
         \includegraphics[width=\textwidth]{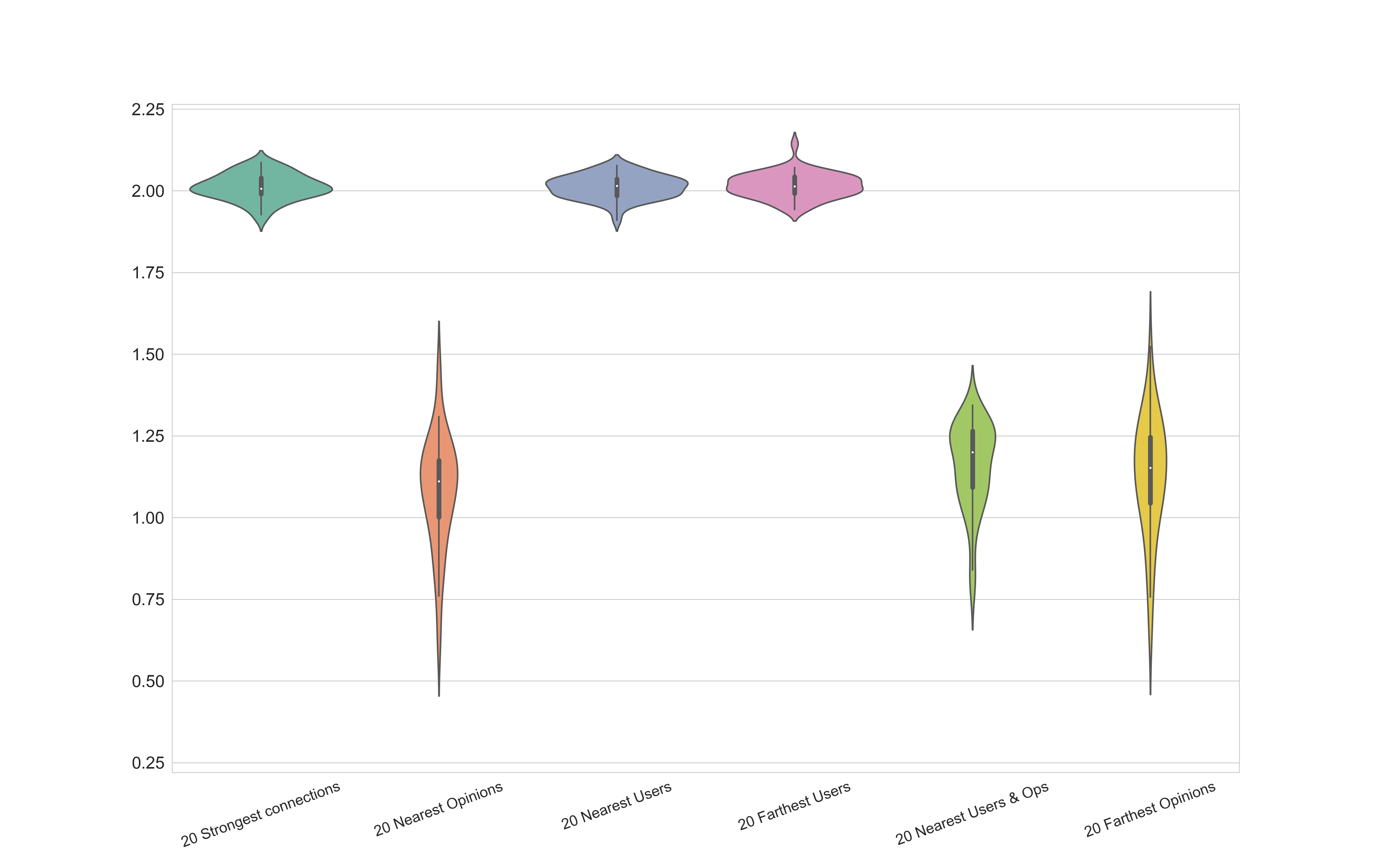}
         \caption{High homophily, low high attention to novelty and initial weights follow power law distribution}
         \label{fig:std_violin_h3a01PL}
     \end{subfigure}
     \hfill
     \begin{subfigure}{0.49\textwidth}
         \centering
         \includegraphics[width=\textwidth]{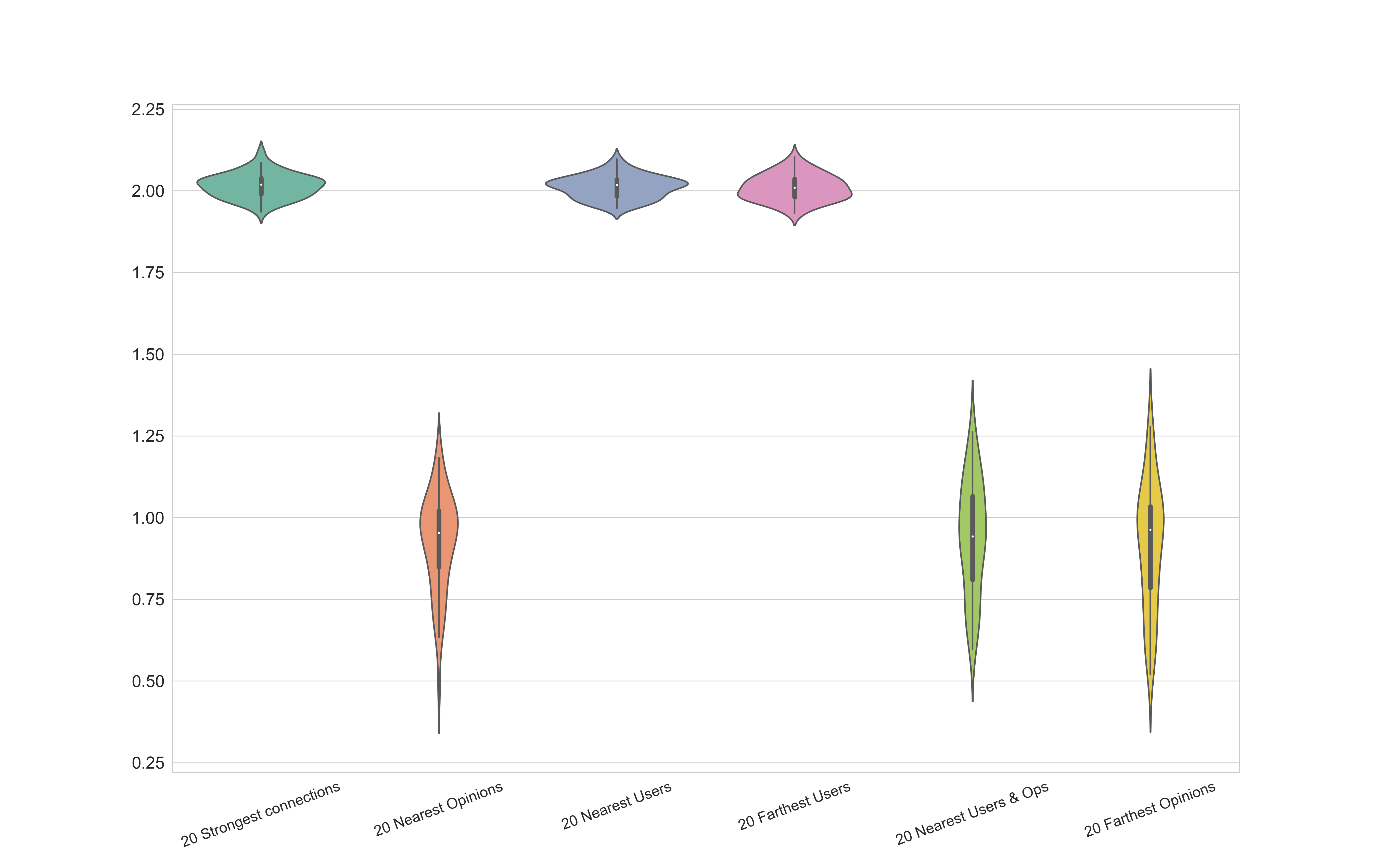}
         \caption{High homophily, low attention to novelty and initial weights follow uniform random distribution}
         \label{fig:std_violin_h3a01UD}
     \end{subfigure}
     \begin{subfigure}{0.49\textwidth}
         \centering
         \includegraphics[width=\textwidth]{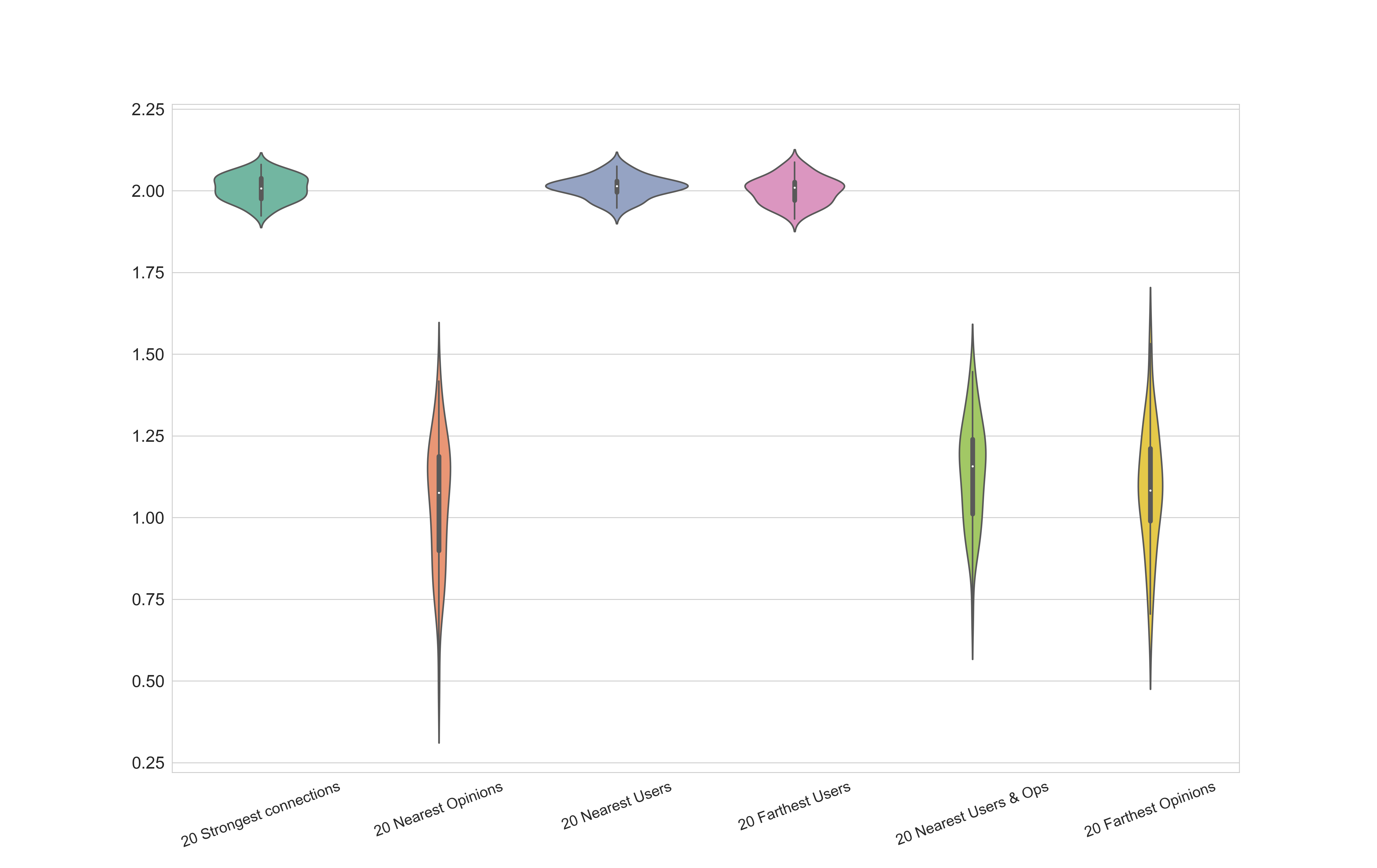}
         \caption{Low homophily, high attention to novelty and initial weights follow power law distribution}
         \label{fig:std_violin_h01a3PL}
     \end{subfigure}
	\hfill
     \begin{subfigure}{0.49\textwidth}
         \centering
         \includegraphics[width=\textwidth]{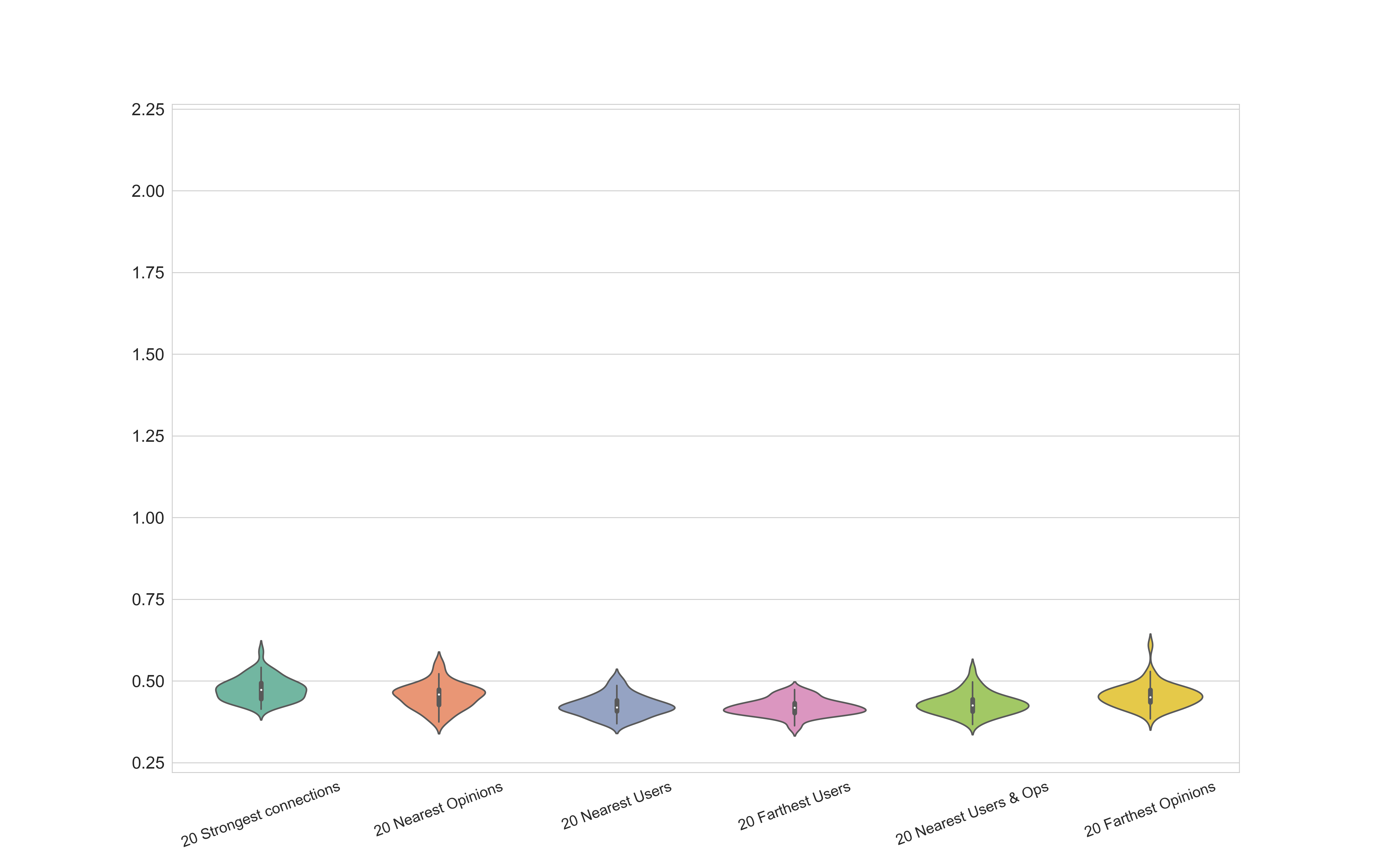}
         \caption{Low homophily, high attention to novelty and initial weights follow uniform random distribution}
         \label{fig:std_violin_h01a3UD}
     \end{subfigure}
     \caption{Violin plots representing distribution of standard deviation of average community idea state for different recommendation strategies and parameter configurations: Each subfigure represents mudularities for the parameter setting described in the caption of subfigure.  In all the cases except (d) when homophily is low, attention to novelty is high and initial edge weights are distributed according to uniform random distribution, opinion recommendation has significantly lower standard deviation representing closeness of different communities in terms of their mean idea state. }
\label{fig: standard_deviation}
 \end{figure}

These findings followed the similar trend as the previous results (\ref{fig:modularity}). We found that opinion recommendation strategies can lead to more ideologically aligned communities with lower standard deviations in terms of their average idea state (Figure \ref{fig: standard_deviation}). This result suggests that these strategies can help bring together users with similar beliefs and preferences, which can have positive effects on social and political cohesion. On the other hand, other recommendation strategies resulted in higher standard deviations, indicating more significant ideological differences between communities. This finding implies that these strategies can result in more ideologically diverse communities, which can be useful in some scenarios, such as in promoting diverse perspectives and viewpoints.
\par The two findings suggest that opinion recommendation strategies can play a crucial role in promoting network and social cohesion by mitigating network fragmentation and bringing together users with similar beliefs and preferences. In contrast, other recommendation strategies can result in more diverse and ideologically different communities, which can be valuable in promoting diverse perspectives and viewpoints. Therefore, it is crucial to carefully consider the intended outcomes and network characteristics when designing and implementing recommendation and filtering strategies. Ultimately, a balanced approach that considers both the promotion of network cohesion and diversity of viewpoints may be most effective in fostering healthy and resilient social and information networks. 

 \begin{figure}

    \begin{subfigure}{\textwidth}
         \centering
         \includegraphics[width=\textwidth]{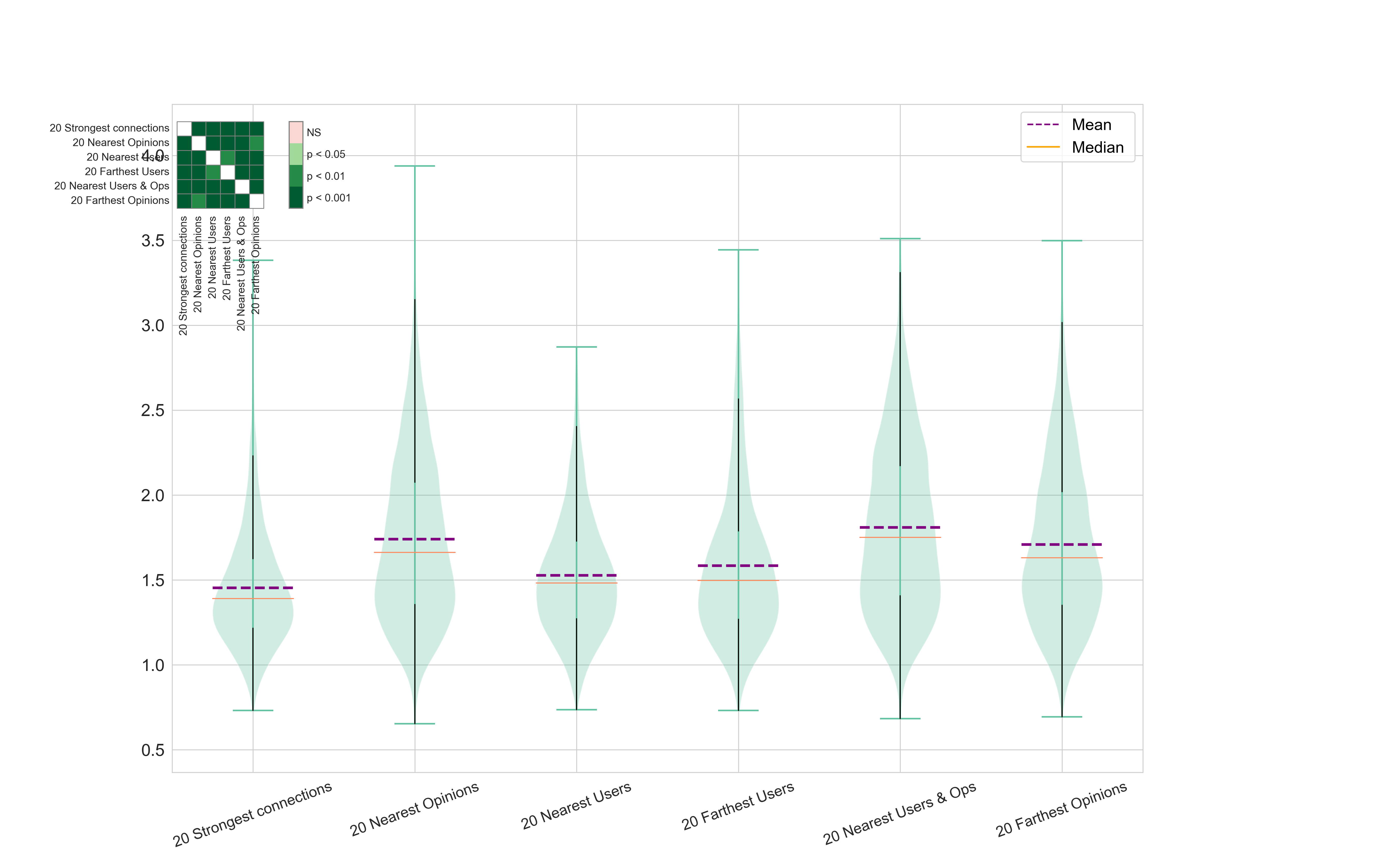}
         \caption{High homophily, low attention to novelty and initial weights follow power law distribution}
         \label{fig:ecc_violin_h3a01PL}
     \end{subfigure}
     \hfill
     \begin{subfigure}{\textwidth}
         \centering
         \includegraphics[width=\textwidth]{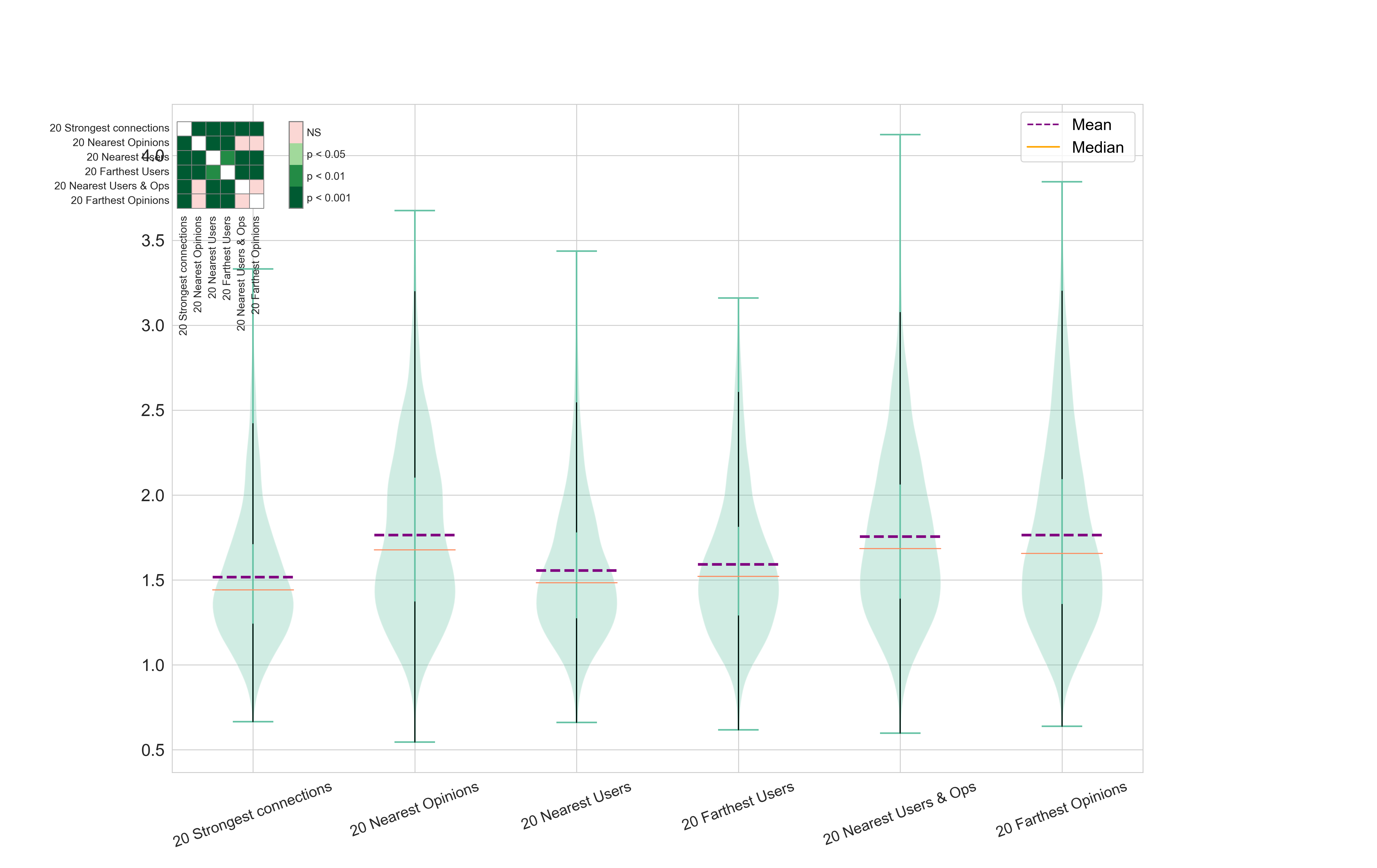}
         \caption{High homophily, low attention to novelty and initial weights follow uniform random distribution}
         \label{fig:ecc_violin_h3a01UD}
     \end{subfigure}
     \end{figure}
     \begin{figure}[H]\ContinuedFloat
     \begin{subfigure}{\textwidth}
         \centering
         \includegraphics[width=\textwidth]{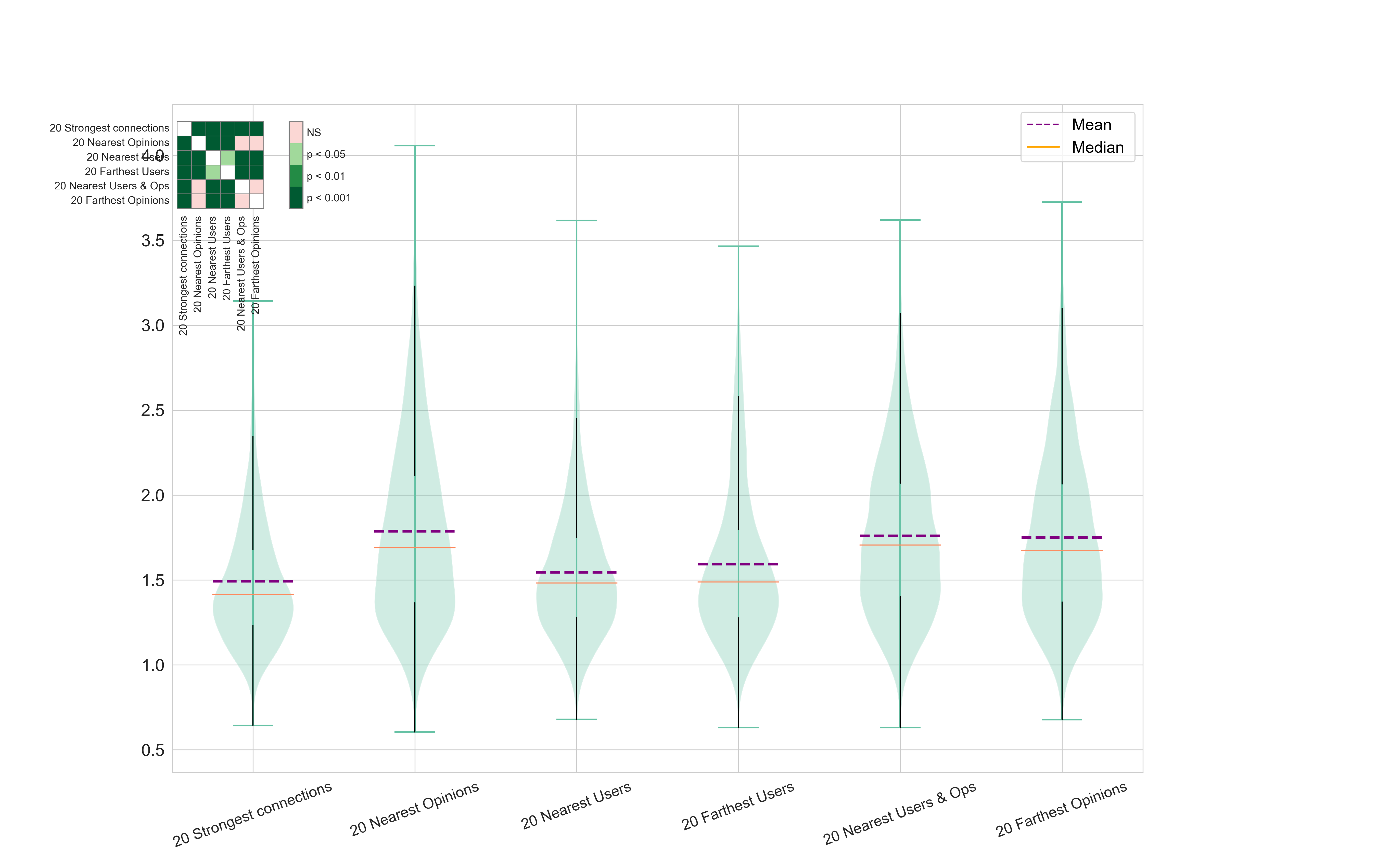}
         \caption{Low homophily, high attention to novelty and initial weights follow power law distribution}
         \label{fig:ecc_violin_h01a3PL}
     \end{subfigure}
     \begin{subfigure}{\textwidth}
         \centering
         \includegraphics[width=\textwidth]{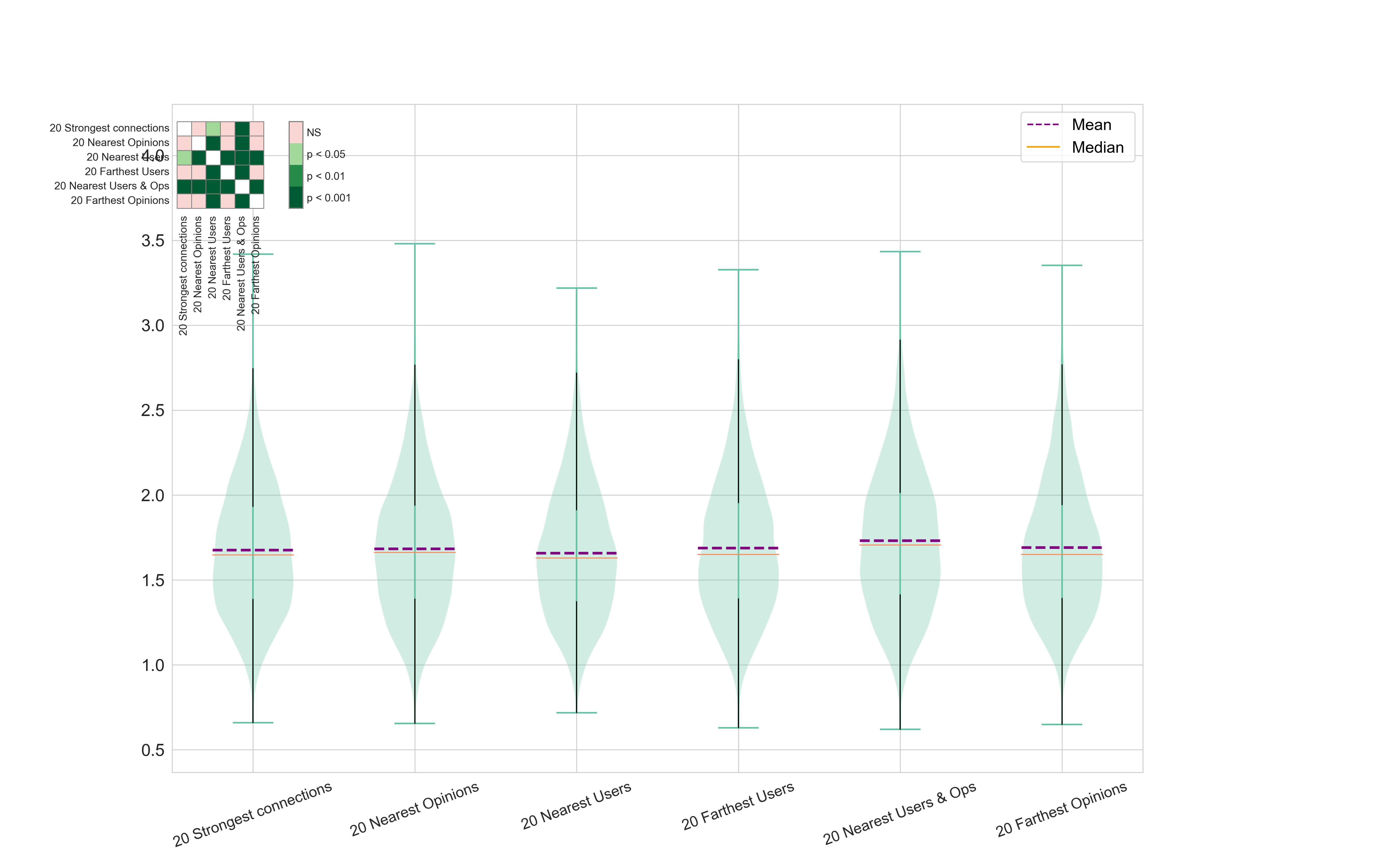}
         \caption{Low homophily, high attention to novelty and initial weights follow uniform random distribution}
         \label{fig:ecc_violin_h01a3UD}
     \end{subfigure}
     \caption{Violin plots representing distribution of opinion eccentricity for different recommendation strategies: Each subfigure represents distribution of eccentricity for the parameter setting described in the caption of the subfigure. Except the case of low homophily and initial weights following uniform random distribution, eccentricity is significantly high when opinion recommendation strategy is employed. Heatmap in each subfigure shows the significance level of difference.}
\label{fig:reco_eccentricity}
 \end{figure}
 \subsection{Effect of recommendation strategies on idea generation process}
Our study on the impact of personal behavioral traits on the idea generation process in the presence of different recommendation strategies has yielded interesting insights. One of our key findings is that opinion recommendation strategies lead to significantly higher concentration of eccentric individual opinions compared to baseline (SC) and user recommendation strategies (NU, FU) (Figure \ref{fig:reco_eccentricity}). This behavior can be explained with the concept of knowledgebase and definition of eccentricity provided in the work by Pandey et al\cite {pandey2023}. Opinion recommendation will increase the diversity of a user`s knowledgebase and hence a newly generated idea will have greater distance from the center of knowledgebase.  It is important to note that eccentricity is measured in the context of the neighbors and community a user is part of, and it may not reflect global eccentricity. In other words, an opinion that may seem eccentric within a tightly knit community may not be eccentric globally when viewed in the context of the entire network. Therefore, it is essential to consider the network structure and community dynamics when analyzing the eccentricity of individual opinions. This finding is particularly significant because it highlights the potential of opinion recommendation strategies to promote diversity of thought and ideas from a user's perspective, which is crucial for innovation and progress in any field. However, in the case of user recommendation strategies, where individuals with similar idea states form strong communities, there is less space for a user to present a diverse and eccentric opinion in the neighborhood. This observation underscores the risk of echo chambers and filter bubbles, where users only interact with like-minded individuals and miss out on exposure to a variety of viewpoints.

\section{Conclusion}
In summary, our study focused on understanding the impact of different recommendation strategies on the idea generation process and network cohesion. Our first finding was that the use of opinion recommendation strategies consistently discouraged network fragmentation across most scenarios. We found that when users are exposed to diverse opinions and perspectives, they are more likely to maintain connections with others, contributing to a more cohesive network. This result highlights the potential of these strategies to mitigate the negative effects of echo chambers and filter bubbles, which can lead to greater ideological polarization and network fragmentation. These findings were supported by the observation we made in terms of standard deviation of average idea states of communities. We saw that opinion recommendation strategies can lead to more ideologically aligned communities with lower standard deviations in terms of their average idea state. This result suggests that these strategies can bring together users with similar beliefs and preferences, which can have positive effects on social and political cohesion. On the other hand, other recommendation strategies resulted in higher standard deviations, indicating more significant ideological differences between communities.	

We also examined the impact of recommendation strategies on the idea generation process and found that in case of opinion recommendation strategies, the eccentricities of opinions are significantly higher than the baseline and user recommendation strategies. This result suggests that in case of opinion recommendations, individual's  knowledgebase gets diversified with recommended opinions and from the individual's perspective, there is more space to express off center opinions, while in case of user recommendation strategies, it is opposite. However, it is essential to note that the eccentricity is measured in the context of the neighbors and user's social neighborhood, and it is not necessary that if an opinion is eccentric within a community, it will be eccentric globally.

Overall, our study provides insights into the impact of different recommendation strategies on network cohesion and idea generation processes. These findings can inform the design of recommendation and filtering algorithms to promote network cohesion and mitigate the negative effects of filter bubbles and echo chambers.

\bibliographystyle{unsrt} 
\bibliography{references}
\end{document}